\documentstyle[prl,aps,epsfig,eqsecnum,twocolumn,subfigure]{revtex}

\catcode`\@=11

\def\maketitle2{\par 
\begingroup
\let\cite\@bylinecite
\def\thefootnote{\fnsymbol{footnote}}%
\twocolumn[\@maketitle2\vskip2pc]%
\thispagestyle{plain}\@thanks
\endgroup
\def\thefootnote{\arabic{footnote}}%
\setcounter{footnote}{0}%
\let\maketitle2\relax \let\@maketitle2\relax
\let\@thanks\relax \let\@authoraddress\relax \let\@title\relax
\let\@date\relax \let\thanks\relax \let\@abstract\relax 
\let\@pacs\relax}

\def\abstract#1{\gdef\@abstract{{\par 
\bgroup
\ifdim\prevdepth=-1000pt \prevdepth0pt\fi
\hsize\columnwidth
\dimen0=-\prevdepth \advance\dimen0 by17.5pt \nointerlineskip
\small\vrule width 0pt height\dimen0 \relax}{~~}#1\egroup}}

\def\pacs#1{\gdef\@pacs{{\par 
\bgroup
\hsize\columnwidth \parindent0pt
\ifdim\prevdepth=-1000pt \prevdepth0pt\fi
\dimen0=-\prevdepth \advance\dimen0 by20pt\nointerlineskip
\egroup} PACS numbers:~#1}}

\def\preprint#1{\gdef\@preprint{Preprint number:~#1}}

\def\@maketitle2{
\@title
\ifdim\prevdepth=-1000pt \prevdepth0pt\fi
\@authoraddress
\@date
\begin{list}{}{\leftmargin=0.10753\textwidth \rightmargin=\leftmargin
\itemsep=1pc\partopsep=-1pc}
\item\@abstract
\item\@pacs; \@preprint

\end{list}
}

\catcode`\@=12

\renewcommand{\thefootnote}{\fnsymbol{footnote}}

\newcommand {\tom}{{\tilde \omega}}
\newcommand {\tOm}{{\tilde \Omega}}

\newcommand {\ts} {{\tilde \sigma}}
\newcommand {\tsd} {{\dot{\tilde \sigma}}}
\begin{document}
\title{Dynamics of the chiral phase transition at finite chemical potential}
\author{Wenjin Mao}
\address{Department of Physics, Boston College, Chestnut Hill, MA
02167}
\author{Fred Cooper and
Anupam Singh}
\address{Theoretical Division, MS B285, Los Alamos National
Laboratory, Los Alamos, NM 87545}
\author{Alan Chodos}
\address{Department of
Physics, Yale University, New Haven, CT 06520-8120. \\
Present Address: American Physical Society, One Physice Ellipse, College
Park, MD 20740.}
\date{\today}
\preprint{LA-UR-00-3104}
\abstract{We study the dynamics of the chiral phase transition
at finite chemical potential in the Gross-Neveu model in the leading order in large-$N$ approximation.
We consider evolutions starting in local thermal and chemical equilibrium
in the massless unbroken phase for conditions pertaining to traversing
a first or second order phase transition.   We assume boost invariant
kinematics and determine the evolution of 
the order parameter $\sigma$, the energy density and pressure  as well as the
effective temperature, chemical potential and interpolating number densities
as a function of $\tau$. }
\maketitle2 

\begin{narrowtext}
The phase structure of QCD at non-zero  temperature and baryon density
is important for the physics of neutron stars and relativistic heavy ion
collisions.  The phase structure for two massless quarks
\cite{ref:review} reveals a rich structure. At low temperature and chemical 
potential, the ground state has broken chiral symmetry. At higher chemical
potential one finds a superconducting phase. The transition out of the 
chirally broken phase as one increases the temperature is second order at 
low chemical potential and then changes to first order as we increase
the chemical potential \cite{tricritical}. 

Recently we found a simple model which has a similar phase
structure \cite{ref:us} to that described above, i.e.
both chiral and superconducting transitions as well as asymptotic freedom. 
Here we consider a special limit without a superconducting phase, where
the  model reduces to the  Gross-Neveu (GN) model \cite{ref:GN} whose
Lagrangian is
\begin{equation} 
 {\cal L} = - i \bar {\Psi}_i \gamma^{\mu}
\partial_{\mu} \Psi_i - {1 \over 2} g^2 \left(\bar {\Psi}_i\Psi^i \right)^2,
\end{equation}
which is invariant under the discrete chiral group: $\Psi_i \rightarrow
\gamma_5 \Psi_i$. 
In leading order in large $N$ the effective action is
\begin{equation}
S_{eff}= \int d^2 x \left[ -i 
\bar {\Psi}_i \left( \not\partial + \sigma \right)\Psi^i -
\frac {\sigma^2 }{2 g^2 } \right] + {\rm tr}{\rm ln} S^{-1}[\sigma],
\label{boost_Sf}
\end{equation}
where $ S^{-1}(x,y)[\sigma]  = \left(\gamma^{\mu} \partial_{\mu} + \sigma
\right) \delta (x-y)$.

The phase structure of the GN model at finite temperature
and chemical potential  in this approximation has
been known for a long time \cite{ref:us} \cite{ref:GN2}and is summarized
in Fig. 1. 

\begin{figure}
   \centering
   \epsfig{figure=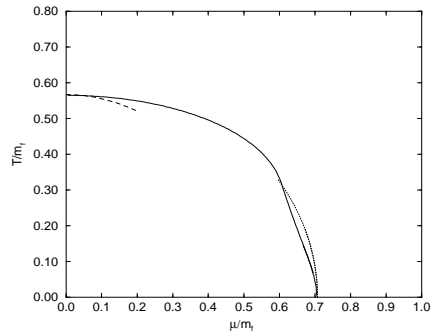,width=2.5in,height=2.in}
\caption{Phase structure at finite temperature and chemical potential $\mu$.
The phase below the line has $ \langle {\bar \psi} \psi \rangle \neq 0$.}
   \label{fig:phase}
\end{figure}

The phase structure is  determined from
 the renormalized effective potential 
\begin{eqnarray}
&& V_{eff} (\sigma^2, T, \mu) = {\sigma^2 \over 4 \pi} ~
[\ln~ {\sigma^2 \over m_f^2}
- 1] \nonumber \\
 && -{2 \over \beta}
\int_0^{\infty} {dk
\over 2
\pi} ~[\ln~ (1 + e
^{- \beta (E - \mu)})+ \ln (1 + e^{- \beta (E  + \mu)})]
\end{eqnarray}
Here $m_f$ is the physical mass of the
 fermion in the vacuum sector.
The tricritical point occurs at $
 {\mu_c \over m_f} = .608, $ ${T_c  \over m_f} = .318.$
We have chosen to renormalize the
effective potential so its value  at $T=0$ in the {\it false} vacuum 
$\sigma=0$ is zero.  In the {\it true} vacuum $\sigma= m_f$ the energy density
has the value $ \epsilon/m_f^2 = - \frac{1}{4 \pi}.$  

  Following a heavy ion collision, the ensuing plasma expands and cools
traversing the chiral phase transition.
In hydrodynamic simulations of these collisions,
a reasonable approximation is to treat the expansion 
as a 1+1 dimensional boost invariant expansion\cite{Bjorken,PRD}
along  the beam ($z$) axis.
In this approximation, the  fluid velocity scales as $z/t$. In 
terms of the variables fluid rapidity $\eta = {1 \over 2} ~{\rm ln }~\left({t+z
\over t-z} \right)$ and fluid proper time  $\tau = (t^2-z^2) ^{1/2}$, 
physical quantities such as $\sigma,\epsilon$  become independent of $\eta$,
as discussed in refs. 
\cite{Bjorken,PRD} and applied to the problem of disoriented
chiral  condensates in ref.\cite{ref:DCC}.
We note that related nonequilibrium techniques have also been developed 
in ref. \cite{bdhnoneq} and applied to the problem of  disoriented
chiral  condensates in ref. \cite{dccbdh}. 
 Although the
effective mass $\sigma$
is a function solely of $\tau$, two-point correlation functions depend on 
$\eta$ as well. 

We shall use the metric convention $(-,+)$. 
In our approximation, the dynamics are described
by the Dirac equation with
self-consistently determined  mass term.  
Rescaling the fermion field,
$ \psi(x)=  {1 \over \sqrt{\tau}} \Phi(x),$ and introducing conformal time $u$
via  $\tau= {e^u \over m}$,
we obtain
\begin{equation}
\left[ \gamma^0 \partial_ u
+ \gamma^3 \partial_\eta + \tilde{\sigma}  (u) \right] \Phi(x)  =0\,,
\label{boost_Dirac3}
\end{equation}
where
$\tilde{\sigma}  (u) = \sigma \tau = {\sigma  \over m}  e^u$ and
$\tilde{m_f}  = m_f \tau .$

Further letting $ g^2 =\lambda/2N$ we have the gap equation
\begin{equation}
\sigma =- i {\frac{\lambda}{2 N}}\left \langle \left[ \Psi^{\dag}_i, {\gamma}
^{0} \Psi_i \right]\right \rangle \equiv  - i {\frac{\lambda}{2 }}\left
\langle \left[ \psi^{\dag}, {\gamma} ^{0} \psi
\right]\right \rangle  ,
\label{gap}
\end{equation}
where we have assumed $N$ identical $\Psi_i = \psi$.

These equations are to be solved subject to initial conditions at
$\tau = \tau_{0}$. It is sufficient to describe 
the initial state of the charged fermion field  by the initial particle and
anti-particle number
densities, which we take to be  Fermi-Dirac
 distributions described by  $\mu_0$ and
 $T_0$.

Expanding the fermion fields $\Phi$  in terms of Fourier modes at
fixed
conformal time $u$,
\begin{equation}
\Phi (x) = \int {d k_{\eta} \over 2 \pi}\{b({k})
\phi^{+}_k(u)
 e^{i k_{\eta} \eta}
+d^{\dagger}({{-k}}) \phi^{-}_{{{-k}}}(u)
e^{-i k_{\eta} \eta}   \},
\label{boost_fieldD}
\end{equation}
the $\phi^{\pm}_{{ k}}$ then obey
\begin{equation}
 \left[\gamma^{0} {d\over d u }
+i {\gamma^{3}} k _{\eta}
 + \tilde{\sigma}(u) \right]\phi^{\pm}_{{k}}(u) = 0.
\label{boost_mode_eq_D}
\end{equation}
 The superscript $\pm$ refers to positive- or
negative-energy solutions. 
Introducing mode functions $\phi^{\pm}_{{ k}}(u)$ via
\begin{equation}
\phi^{\pm}_{{ k}}(u) =  \left[-\gamma^0 {d\over d u}
- i {\gamma^{3}} k_{\eta}
 + {\tilde\sigma}(u ) \right] f ^{\pm}_{k}(\tau) \chi^{\pm} ,
\label{boost_mode_eq_f}
\end{equation}
where the momentum independent spinors $\chi^{\pm} $ are chosen to be
the orthornomal $\pm 1$ eigenstates of $i\gamma^0$,
we obtain the second order equations:
\begin{equation}
\left(- \frac{d^2}{d u^2}-
\tom_k^2  \pm  i~~{d \ts \over d u}  \right )
f^{\pm}_{k}(u) = 0,
\label{eq:mode}
\end{equation}
where 
 $\tom_k^2= k_{\eta}^2 +\ts^2(u).$ 
We parameterize the positive-energy solutions $f^{+}_{k}$
in a similar manner to  Eq.~(3.1) of Ref.~\cite{WKB}:
\begin{eqnarray}
f_{k}^+ (u) = 
 \frac {N_k}{\sqrt{2\tOm_{k}(u)}} \exp\left \{ \int_{0}^{u}
\left ( -i\tOm_{k} (u')
- \frac {{\dot{\ts}
}(u')}
{2\tOm_{k} (u')}
\right )
du'\right \}  .
\label{boost_ansatz_D} \nonumber 
\end{eqnarray}

Using eqs.(\ref{boost_fieldD}, \ref{boost_mode_eq_f})and the definitions:
$\langle b^{\dag}(k) b(q) \rangle= 2 \pi \delta(k-q) N_+(q)$ and
$\langle d^{\dag}(k) d(q) \rangle = 2 \pi \delta(k-q) N_{-}(q)$,
we obtain for the gap equation
\begin{eqnarray}
\tilde{\sigma}
&& = \lambda    \int {dk_{\eta}  \over 2 \pi}(1- N_+(k) - N_{-}(k)) R_k(u) \label{eq:sigdim}
\end{eqnarray}
where $R_k(u)= 1 - 2 k_{\eta}^2 ~~|f^+_k (u)|^2. $
and 
\[
 \lambda^{-1} =  \int {dk_{\eta} \over 2 \pi}
 \frac{1}{\sqrt{k_\eta^2 + \tilde{m}_f^2}}= \int {dk \over 2 \pi}
\frac{1}{\sqrt{k^2+m_f^2} }. 
\]
This equation is solved simultaneously with eq. (\ref{eq:mode}).

We take our initial state to be 
in local equilibrium so that 
$N_{\pm}(k,\mu,T) = [e^{( \omega_k(0)\mp \mu) /T}+1 ]^{-1}$ 
where
$\omega_k(0) = E$  $ = \sqrt{k^2+\sigma^2(0)}$ $= { \tom_k(0) \over \tau_0}.$
Since we start our simulation in the unbroken mode,
$ \ts(0) =0.$
We choose the initial $ \tau_0 ={1 \over m_f}$ and measure the
proper time in these units.
We use adiabatic initial conditions on the mode functions $f$, i.e.
$f_k(0) = {N_k \over \sqrt{2 \tom_k}}$, 
$\dot{f}_{k}^+ (0) = - i  \tom_k   f_{k}^+ (0)$ and
$N_k^2 = [\tom_k(0) +\ts(0)]^{-1}.$ To obtain non-trivial dynamics in this
mean field approximation at high temperatures, it is necessary to explicitly
break the chiral symmetry by giving $\tsd$ a small initial value which we
choose to be $\tsd(0)=
10^{-3}$. 

We have studied three separate starting points on
the phase diagram of Fig. 1 in our
numerical simulations. We  determined the energy density and the pressure
from the expectation value of the energy momentum tensor as described in 
\cite{PRD}.
In the $\eta$, $\tau$ coordinate system $T_{\mu \nu}$  is diagonal
which allows us to read off the comoving pressure and energy density.
After renormalization we obtain
\begin{eqnarray}
&&\epsilon(\tau) \tau^2 = 
\int_{0}^{\tilde {\Lambda}} {dk_\eta \over 2 \pi} \biggl [ {\tilde{\sigma}^2
\over  \sqrt{k_{\eta}^2 + \tilde{m}_f^2}} + 4
\Omega_k (\tilde{\sigma}^2 - \omega_k^2) |f_k|^2 \biggr .  \nonumber \\
&&  ( N_{+}+N_{-} )
\biggl . \left[ 
2 \tilde{\sigma} + 4 \Omega_k( \omega_k^2 - \tilde{\sigma}^2) |f_k|^2 + 2
(k_{\eta} - \tilde{\sigma})\right] \biggr ], \label{eq:eps}
\end{eqnarray}
\begin{eqnarray}
p \tau^2 &&= \int_{0}^{{\tilde \Lambda}} {dk_{\eta} \over 2 \pi} 
\biggl [ (1-N_{+}-N_{-}) ~4~(\tilde{\sigma}+ \Omega_k) ({\tilde
\sigma}^2 -\omega_k^2 ) |f_k|^2  \biggr . \nonumber \\
&& \left. + 2 {k_{\eta}^2 \over \sqrt{k_\eta^2 + \tilde{\sigma}^2}} + 2
\sqrt{k_\eta^2 +  \tilde{\sigma}^2} - 2 k_{\eta} -{\sigma^2 \over 
\sqrt{k_{\eta}^2 + \tilde{m}_f^2}}\right]. \nonumber 
\end{eqnarray}
The integrations involve a moving cutoff $\tilde{\Lambda} = \Lambda
\tau$ when the mode functions are truncated at physical $k_z= \Lambda$.  
In the massless phase, one finds that the exact equation of
state is $p=\epsilon.$ To compare our field theory calculation
with a  local equilibrium hydrodynamical model we assume
\begin{equation}
T^{\alpha \beta} = p g ^{\alpha \beta} + (\epsilon+ p ) u^\alpha u^\beta
\end{equation}
The conservation law of energy and momentum
$ T^{\alpha \beta}~ _{;\beta} = 0,$
combined with scaling law
$ v= z/t $ and $p = \epsilon$ yields \cite{Bjorken} 
${\epsilon \over \epsilon_0} = ({\tau_0 \over \tau}) ^{2}$,${T \over T_0} =
({\tau_0 \over \tau})$. 
From Eq. (\ref{eq:eps}) we can also determine 
$p(\mu,T)$ and  $\epsilon(\mu,T)$. Assuming
$ T/T_0 = \tau_0 / \tau$ and 
$\mu /\mu_0 =
\tau_0 /
\tau$ we find that the local equilibrium expressions for $\epsilon$ and $p$
evolve identically to the numerically determined field theory evolution before
the phase transition. 
(We note that in thermodynamic equilibrium
$ dT/T = d\mu /\mu $ \cite{LandauL} and so close to equilibrium we 
expect the temperature and chemical potential to have a similar falloff with 
time. In fact, a different falloff for the two quantities as a function 
of time 
can be viewed as a departure from local thermal and chemical equilibrium.)
With the same assumptions we find
the distributions for  $N_{\pm}$ plotted against $k_{\eta}$ are independent of
$\tau$. This also agrees with the exact evolution before the phase
transition.

We want to understand how the particle number
distributions evolve in time. In relativistic quantum mechanics, particle
number is not conserved. However in a mean field approximation one can
define an interpolating number operator which at late times becomes the outstate
number operator. By fitting the 
interpolating number densities for both fermions and antifermions 
to  Fermi-Dirac distributions \cite{ref:Aarts} we extract the best value of
$\mu$ and $T$ for that value of the proper time. To define  the interpolating
number operator we use a set of orthonormal mode functions $y_k$ \cite{WKB}
which are the adiabatic  approximation to the exact mode functions:
$y^+_k = u_k e^{-i \int~ \tom_k du}$;~~~ $y^-_{k} = v_{k} e^{i \int~ \tom_k
du}$  with
$u_k = { -i \gamma^{\mu} k_{\mu} +\ts \over \sqrt{2 \tom_k (\tom_k+\ts)}}~~
\chi^+ $;~~ $v_{-k}= { i \gamma^{\mu} k_{\mu} +\ts \over \sqrt{2 \tom_k
(\tom_k+\ts)}}~~ \chi^- .$  
The creation and annihilation operators then
become time dependent and the expansion of the quantum field becomes 
\[
\Phi (x) = \int {d k_{\eta} \over 2 \pi}[a({k},u)
y^{+}_k(u)
+c^{\dagger}(k,u) y^{-}_{k}(u)  ] e^{i k_{\eta} \eta}.
\]
This is an alternative expansion to that found in Eq.(\ref{boost_fieldD}).
The two sets of creation and annihilation operators are related by
a Bogoliubov transformation
$a(k,u) = \alpha_k(u) b(k)$ +$\beta_k^\ast d^\dag(k)$;
$c^\dag(k,u) = $- $\beta_k(u) b(k) $+ $\alpha^\ast_k d^\dag(k).$ 
To ensure that at $u=0$ the two number operators match, one
chooses adiabatic initial conditions:  $\phi_k=y_k$, so that 
$\alpha_k(0) = 1$;$\beta_k (0)=0.$
 The interpolating number
operators for fermions and anti-fermions are defined  by
$
N^+(k,u) = \langle  a^{\dag}(k,u)  a(k,u) \rangle ;$~~
$N^-(k,u) = \langle  c^{\dag}(k,u)  c(k,u) \rangle.$
With $
\Delta_k = { \dot{\tOm}_k + \dot{\ts} \over
2 \tOm_k} $  we have explicitly 
\[
|\beta_k|^2 = k_{\eta}^2 { (\tOm_k - \tom_k)^2 + \Delta_k^2 \over 2 \tom_k (\tom_k+\ts)~~~ [\tOm_k^2 +
\tom_k^2 + 2 \tOm_k \ts + \Delta_k^2]},
\]
\[
N^{\pm}(k,u) = N^{\pm}(k) + [1- N^+(k) - N^-(k) ]|\beta_k(u)|^2.
\]

We have solved the simultaneous equations Eqs. (\ref{eq:mode},
\ref{eq:sigdim}) numerically. 
Comparing  $N^{\pm}(k,u)$ with an equilibrium
parameterization we have determined  $T(k,u)$ and $\mu(k,u)$ as a
function of $k$. When these quantities are
independent  of $k_{\eta}= k \tau $ this defines a time evolving
temperature and chemical potential. We found
that $T$ and $\mu$ are independent
of $k$ except at high momentum before the chiral phase
transition. 

From Fig. \ref{fig:tmusig} we see that 
for both the 1st and 2nd order transitions, 
 $\sigma(\tau)$  shows a sharp transition
during evolution from the unbroken mode to the
broken symmetry mode. Before the phase transition the temperature
falls consistent with the equation of state  $p= \epsilon$. 
For the 2nd order transition,  the chemical potential follows
the temperature and falls as $\tau^{-1}$.  After the phase transition,
there is now a mass scale $m_f$ which leads to oscillations of $\sigma$. 
For the 1st order transition the chemical potential falls  
faster than $\tau^{-1}$.  If one traverses the
tricritical regime one finds results for $\mu$ intermediate
between  the two cases displayed.

\begin{figure}
   \centering
  \epsfig{figure=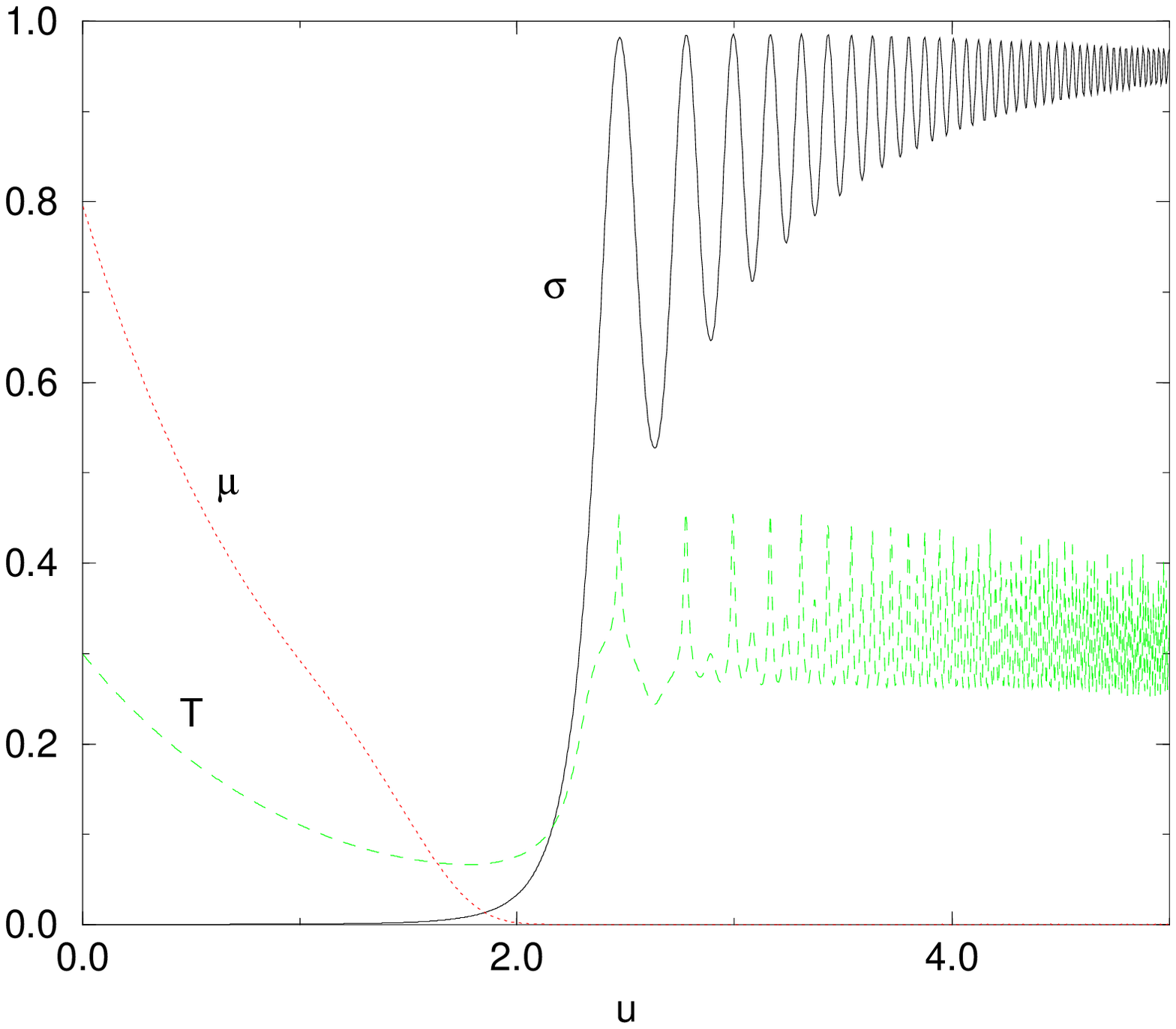,width=2.5in,height=1.8in}
   \epsfig{figure=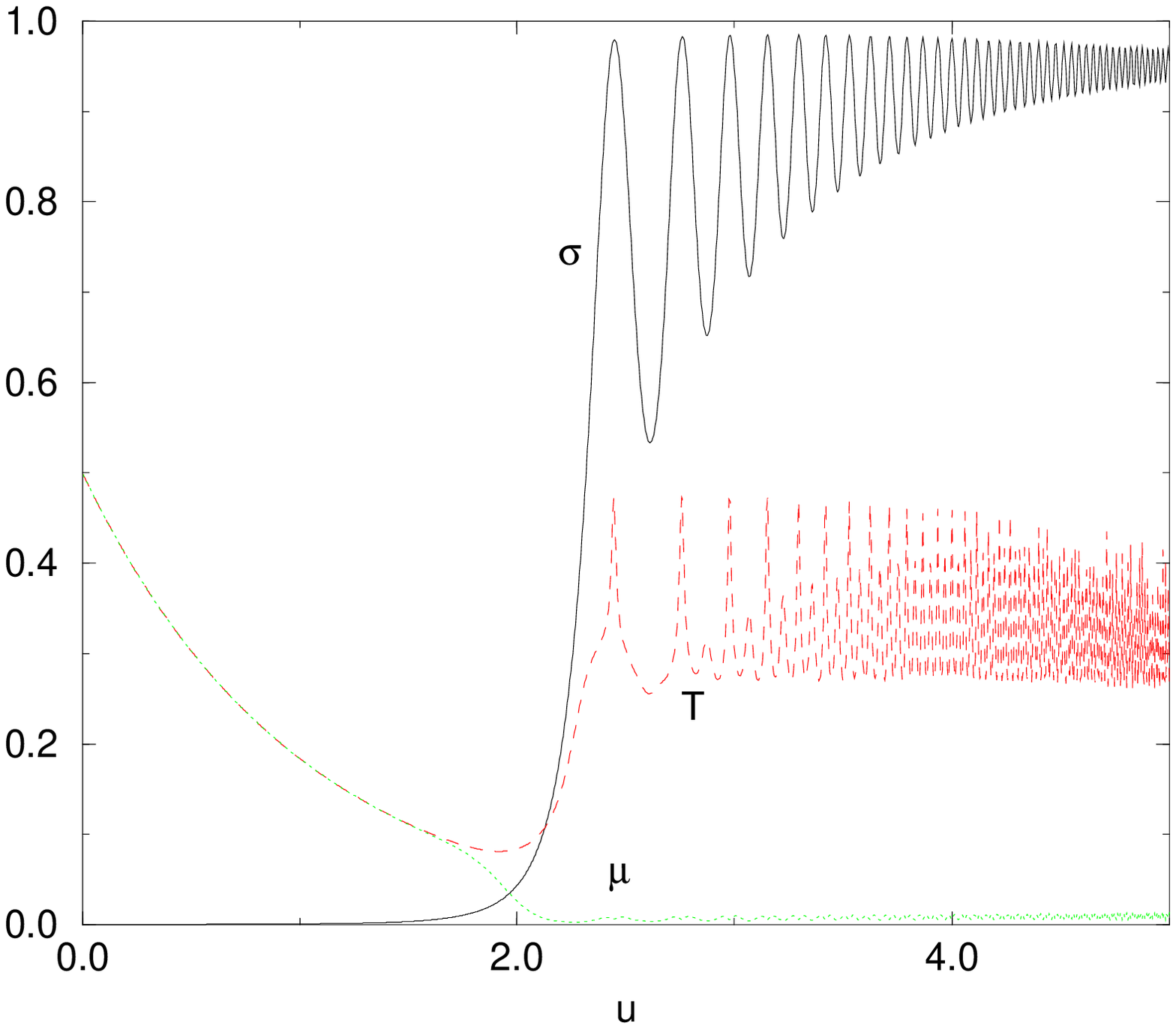,width=2.5in,height=1.8in}
\caption{ Evolution of $T$,$\mu$ and $\sigma$ as a function of $u$.
Top figure is for 1st order transition. Bottom figure is for
2nd order phase transition}    \label{fig:tmusig}
\end{figure}

The order of the transition  has a more noticeable effect on the spectrum of particles and 
antiparticles. If the system evolves in local thermal
equilibrium  with $\sigma=0$, then when $N^\pm (k,u)$ is plotted vs.
$k_\eta= k\tau$ it is independent of $u$. A deviation from this result
is an indication of the system going out of equilibrium.  We  find 
because of the ``latent heat'' released during a first order
transition that the distortion of the spectrum is greatest in that case.
(see Fig. \ref{fig:Nfirstsecond}).
If one traverses the
tricritical regime one finds results intermediate
between  the two cases displayed. 

In local equilibrium with $\sigma=0$, $\epsilon=p \propto \tau^{-2}$ .
Simulations, shown in Fig. \ref{fig:epsp} agree with this
before the  phase transition occurs. 
 After the phase transition we find
that the energy density oscillates around the true broken symmetry value
discussed earlier, namely $  \epsilon_0 = -1/ 4 \pi$. These 
oscillations would be damped if we  included hard scattering
effects \cite{ref:berges}. The details of this calculation as well as a
discussion of correlation functions and the effects of a bare mass will be
presented elsewhere. 

\begin{figure}
   \centering
   \epsfig{figure=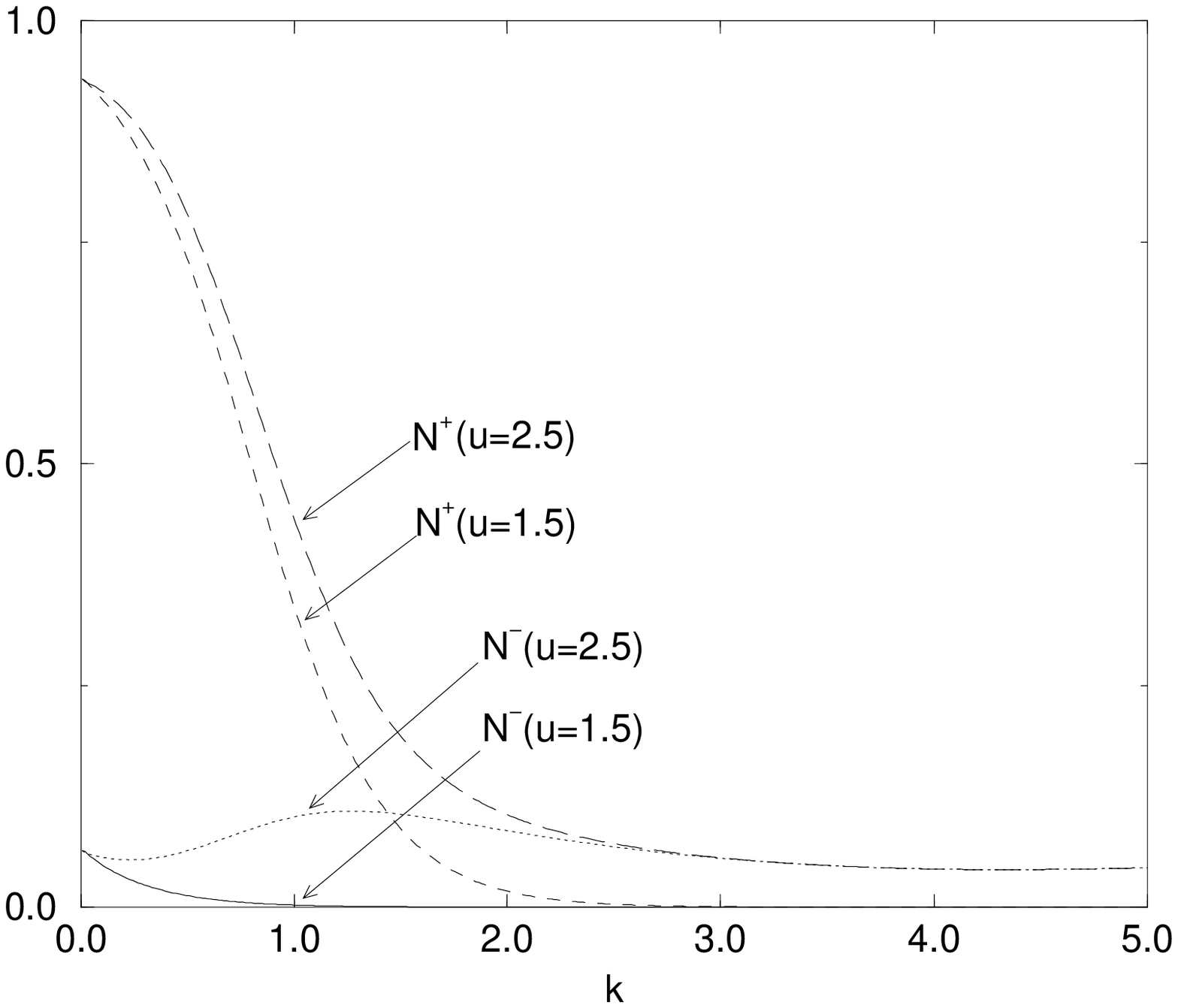,width=2.5in,height=1.8in}
    \epsfig{figure=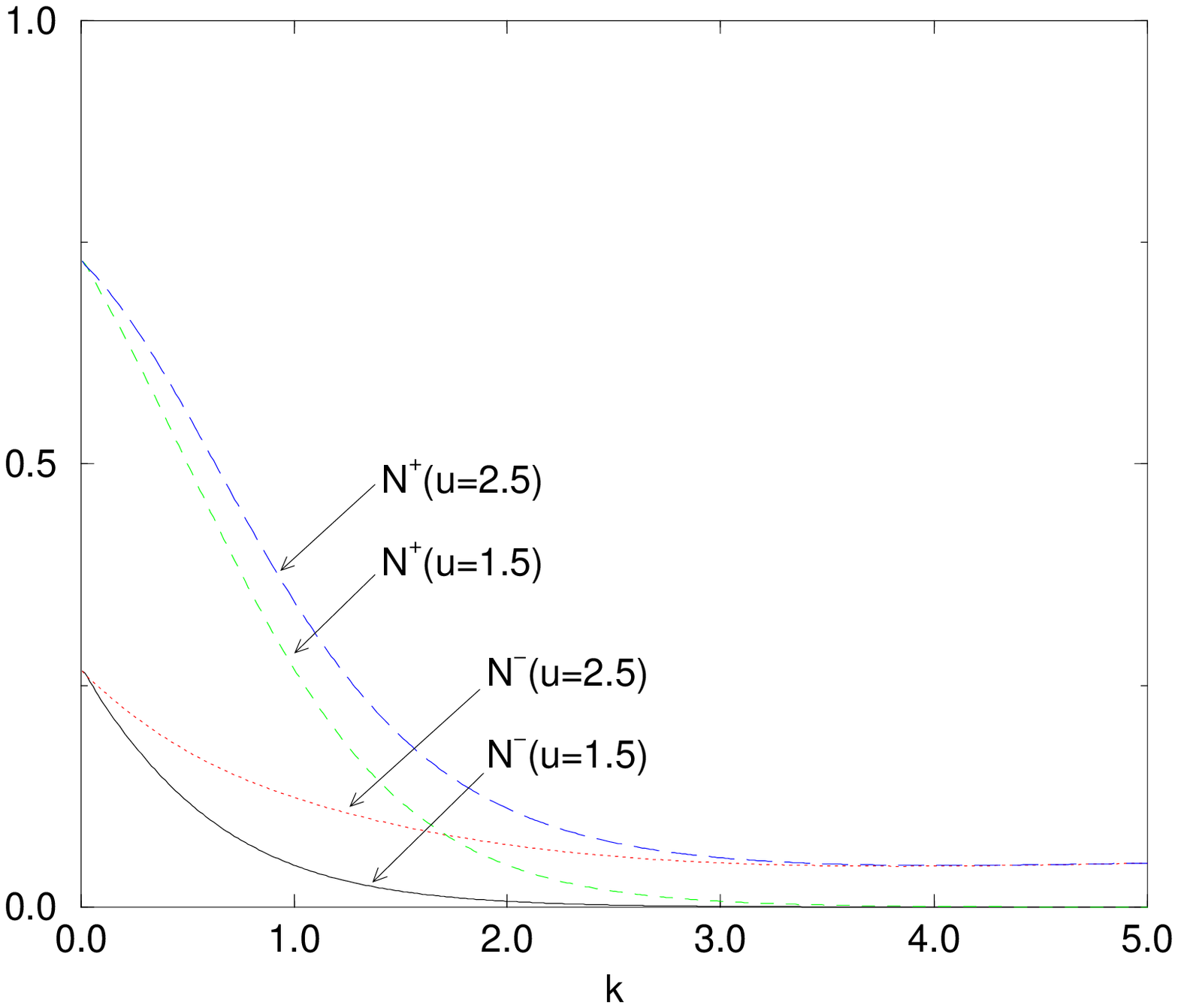,width=2.5in,height=1.8in}
\caption{Evolution of $N_{\pm}$ as a function of $u$.Initial conditions are
same as Fig. \ref{fig:tmusig}.The momentum displayed is $k_\eta= k \tau$}   
\label{fig:Nfirstsecond} \end{figure}  \begin{figure}    \centering
\epsfig{figure=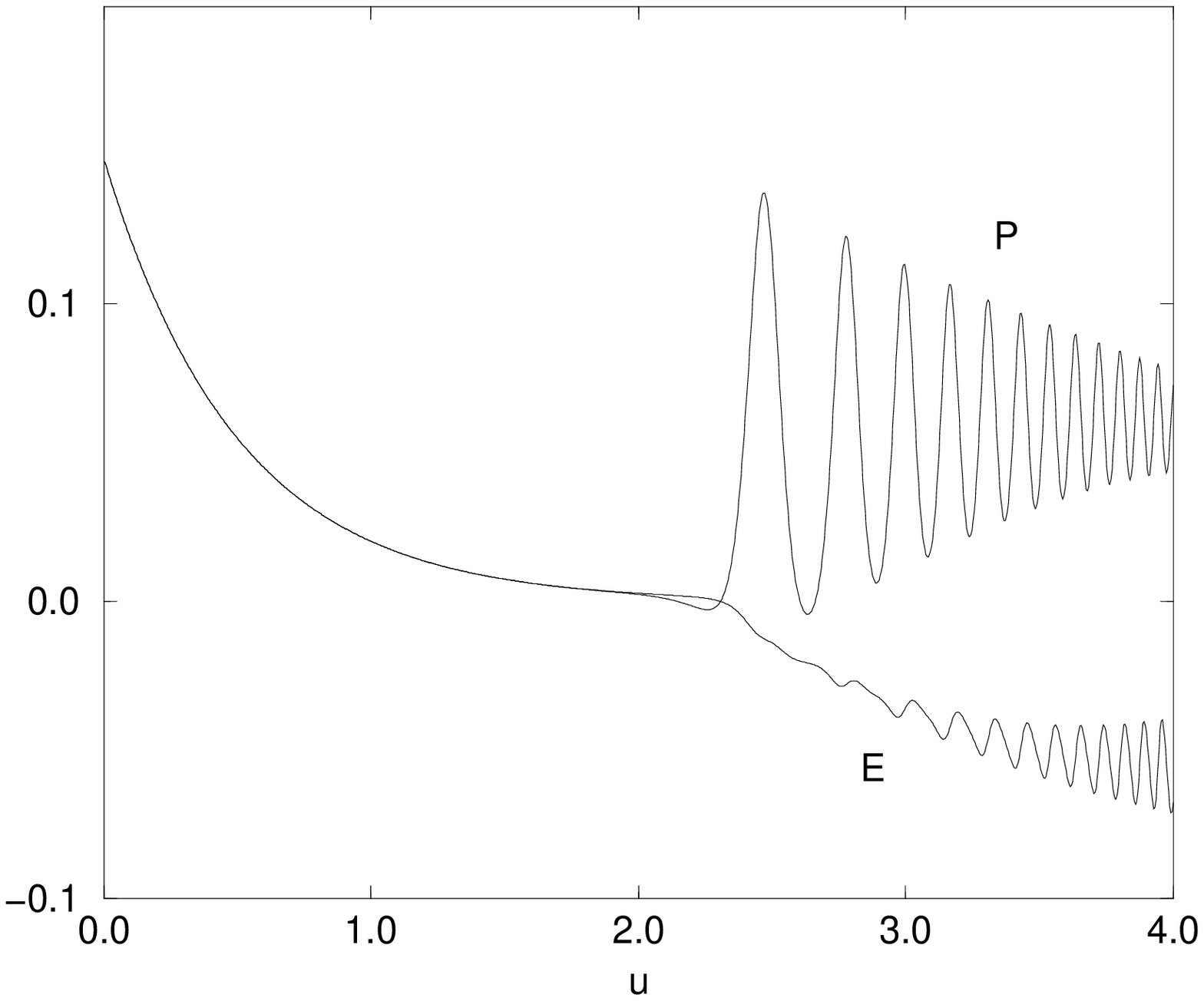,width=2.5in,height=1.8in}
   \epsfig{figure=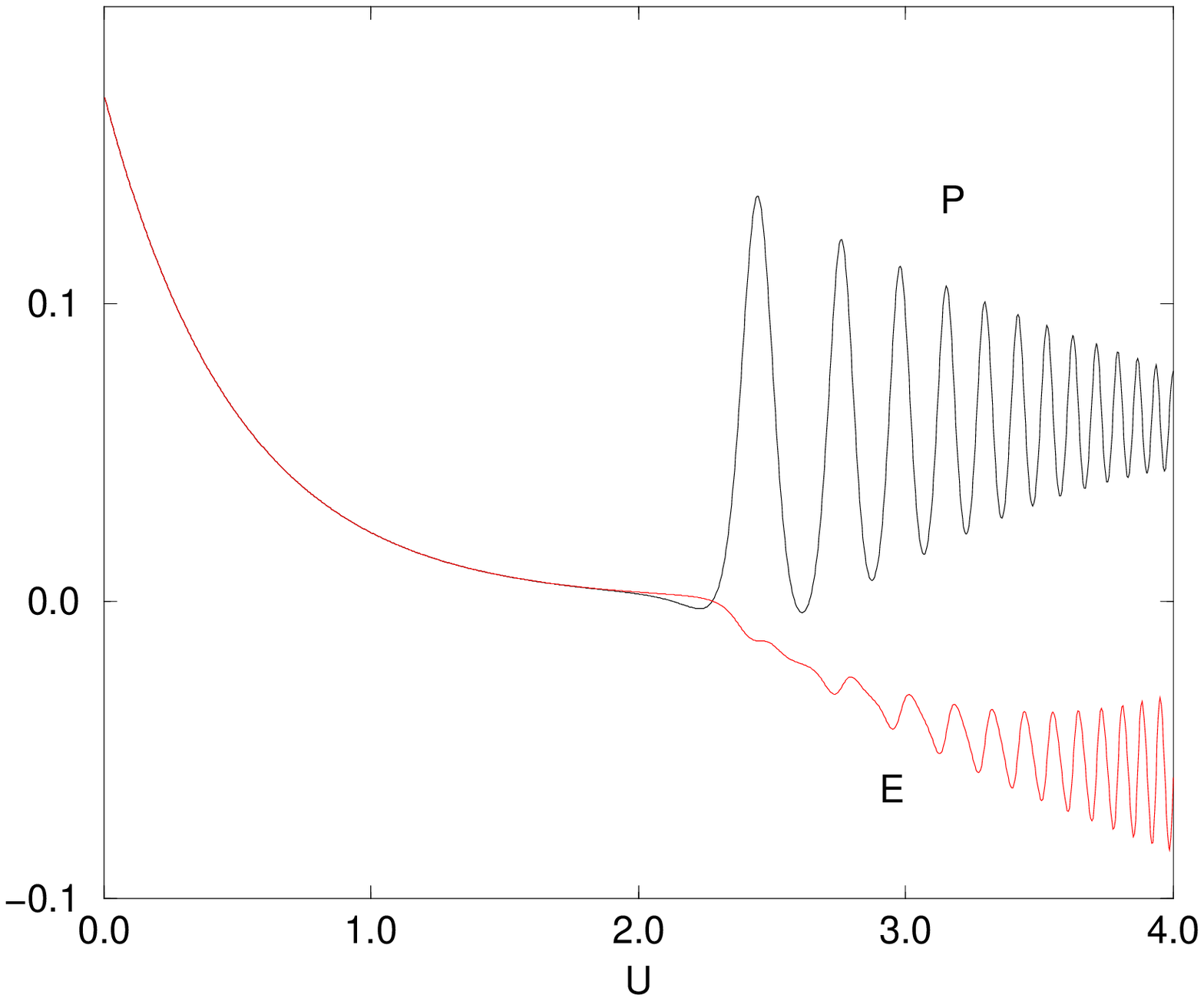,width=2.5in,height=1.8in}
\caption{ Evolution of the pressure  and energy density  as a function of $u$
.Initial conditions are
same as Fig. \ref{fig:tmusig}. }     \label{fig:epsp}
\end{figure}
We would like to thank Emil Mottola, Salman Habib and Dan Boyanovsky for
discussions.

\end{narrowtext}
\end{document}